\documentstyle[12pt]{article}
\def\sab{$S_\alpha^\beta$ }
\def\sabo{$S_\alpha^{\beta}(O)$ }
\def\sa1O{$S_\alpha^{1}({\cal O})$ }
\def\sabk{$S_\alpha^{\beta}(K)$ }
\def\ss{\subset \subset }
\def\TMP{{\it Theor.\ Math.\ Phys.} }

\def\LMP{{\it Lett.\ Math.\ Phys.} }

\def\JMP{{\it J.\ Math.\ Phys.} }
\def\SM{Soloviev,~M.~A., }
\def\plu{plurisubharmonic }

\begin{document}

\thispagestyle{empty}

\baselineskip=0.6cm

\noindent P.~N.~Lebedev Institute Preprint     \hfill
FIAN/TD/2--94\\ I.~E.~Tamm Theory Department       \hfill
\begin{flushright}{March 1994}\end{flushright}

\begin{center}

\vspace{0.5in}

{\Large\bf Towards a Generalized Distribution Formalism for Gauge Quantum
Fields}\footnote{This work was supported in part by a Soros Humanitarian
Foundation Grant awarded by the American Physical Society.  }

\bigskip

\vspace{0.3in}
{\large  M.~A.~Soloviev}\\
\medskip  {\it Department of Theoretical Physics} \\ {\it  P.~N.~Lebedev
Physical Institute} \\ {\it Leninsky prospect, 53, 117 924, Moscow,
Russia}$^{\dagger}$\\

\end{center}

\vspace{1.5cm}

\centerline{\bf ABSTRACT}
\begin{quotation}

We prove that the distributions defined on the Gelfand-Shilov
spaces \sab with $\beta<1$, and hence more singular than hyperfunctions, retain
the angular localizability property. Specifically, they have uniquely
determined
support cones. This result enables one to develop a distribution-theoretic
techniques suitable for the consistent treatment of quantum fields with
arbitrarily singular ultraviolet and infrared behavior. The proof covering the
most general and difficult case $\beta=0$ is based on the use of the theory of
plurisubharmonic functions and H\"{o}rmander's $L^2$--estimates.

\end{quotation}

\vfill

\noindent

$^{\dagger}$ E-mail address: soloviev@td.fian.free.net

\newpage

\setcounter{page}{2}

\section{Introduction}

Recently the problem of an adequate choice of test functions in quantum field
theory \cite{1} came once again into notice with emphasis on infrared
singularities which occur in gauge QFT in connection with the necessary lack of
positivity \cite{2}. Namely, analysis of some explicitly soluble models has
shown [3--5] that these singularities are in general more severe than
those of ultradistributions or even hyperfunctions and that appropriate test
function spaces are \sab by Gelfand and Shilov \cite{6}. The indices $\alpha$
and $\beta$ control respectively the ultraviolet and infrared singularities
whose order increases with decreasing their values, while the Schwartz space
used in the original Wightman formalism can be identified with
$S^\infty_\infty$. Of great importance is the fact that the usual definition of
support has sense only for generalized functions on \sab with $\beta>1$, for
otherwise the configuration-space test functions are analytic. The same is true
referring to the momentum-space test functions and $\alpha>1$. Just then one
uses the term {\it ultradistributions} and the former inequality selects in
fact
the Jaffe strictly localizable fields \cite{7}. The space $S^1_1$, which was
put
forward in \cite{8} as most adequate to Meiman's locality principle \cite{9},
admits a direct generalization of the notion of support and underlies the {\it
Fourier hyperfunction} QFT \cite{10}. As shown in Refs. [1--5], the infrared
behavior of gauge fields treated in generic covariant gauges is so singular
that
one is forced to use spaces with $\alpha<1$ in order to represent the fields as
operator-valued distributions on a Hilbert space. This raises the problem of
formulation of the spectral condition which is certainly similar to that of
generalization of local commutativity beyond the localizability bound.
Fortunately, the latter problem has received an extensive study which
resulted particularly in new proofs of the spin-statistics and {\it TCP}
theorems covering nonlocal quantum fields \cite{11,12}.

In Refs. [13--15], we have shown that the mathematical techniques used in
nonlocal QFT can essentially be improved by adoption some ideas of the theory
of
hyperfunctions. The principal result is evidence of the existence of uniquely
defined support cones which replace supports in the nonlocalizable case. This
enables one to extend a great part of the theory of distributions beyond the
hyperfunctions and makes generalization of the Wightman approach quite natural.
However those proofs are based on exploiting test functions of rapid decrease
from the subspace $S^\beta_{1-\beta}$ and so are inapplicable to $S^0_\alpha$
since $S^0_1$ is trivial. The main aim of this Letter is to fill this gap
because the spaces $S^0_\alpha$ (and $S^\beta_0$) are of special interest and
provide us with the widest distributional framework adequate to quantum
fields with arbitrarily singular infrared (ultraviolet) behavior. There are
three essential steps in proving: The extension of the scale of Gelfand-Shilov
spaces to test functions defined on open and closed cones (Sec.2), the
representation of these spaces in terms of complex variables (Sec.3), and  the
employment of H\"{o}rmander's $L^2$--estimates for solutions of nonhomogeneous
Cauchy-Riemann equations (Sec.5). In the last section 6, we indicate several
immediate applications of the obtained results whose detailed presentation will
be given in forthcoming papers.

It should be remarked that, following H\"{o}rmander, we prefer to use the term
{\it distributions} instead of {\it generalized functions} under arbitrary
singularity and, as usual, we call {\it tempered} the Schwartz distributions
whose order of singularity is finite.

\section{Nonlocalizable distributions versus hyperfunctions}

{\large Definition 1.} {\it Let $O$ be an open set in ${\bf R}^n$. The space
\sabo $\>$ consists of all complex valued infinitely differentiable functions
on
$O$ with the property that the norm

$$
\Vert \varphi \Vert_{O,a,b}\> \stackrel{def}{=}\> \sup_{x\in O, k,
q}\,\frac{\vert x^k\partial^q\varphi (x)\vert}{a^{|k|}b^{|q|}k^{\alpha
k}q^{\beta q}}
         \eqno{(1)}
$$

\noindent
is finite for some positive $a, b$ dependent on $\varphi$. }

Here $k$ and $q$ are multi-indices and the standard notation relating to
functions of several variables is used. The set of norms (1) determines
naturally an inductive limit topology on \sabo whose strong dual space is
denoted by $S_\alpha^{\prime \beta}(O)$. By making certain regularity
assumptions regarding $O$, it is easy to show that the former is a DFS
space and the latter is an FS space, see \cite{14} for details. Hence they have
nice topological properties in perfect analogy to the original spaces \sab =
$S_\alpha^{\beta}({\bf R}^n)$ and $S_\alpha^{\prime \beta}$. When dealing with
hyperfunctions, it is useful to bear in mind the following

{\large Proposition 1.} {\it The inductive limit of the spaces $S^1_\alpha
(|x|\!\!>\!\!\!R) \quad ( R\rightarrow \infty )$ is a Hausdorff space wherein
$S^1_\alpha$ is embedded.}

This is the case due to analyticity of these test functions. It implies that
the
mappings $S^1_\alpha (|x|\!\!>\!\!\!R_1)\rightarrow S^1_\alpha
(|x|\!\!>\!\!\!R_2) \quad (R_1<R_2)$ are injective and hence the limit is also
DFS. Clearly every element of $S^{\prime 1}_\alpha$ which is continuous under
the topology induced on $S^1_\alpha$  by that of the inductive limit should be
considered as attached to infinity, if a localization is possible at all. By
contrast, in the case $\beta>1$ the kernel of the canonical mapping from
\sab into the corresponding inductive limit is everywhere dense and the
only ultradistribution continuous under the induced topology is zero.
Therefore
it is reasonable to adapt the definition of the presheaf {$S_\alpha^{1}(O)$} to
a compactification of ${\bf R}^n$. Following Kawai \cite{16}, we use the radial
compactification  ${\cal R}^n$ and identify the space \sa1O, where ${\cal O}$
is
an open subset of ${\cal R}^n$, with $S_\alpha^{1}({\cal O}\cap R^n)$. Then the
space $S_\alpha^1 ({\cal K})$ corresponding to a compact set ${\cal K} \subset
{\cal R}^n$ is defined by $S_\alpha^1 ({\cal K}) = \hbox{inj\,lim}\> S_\alpha^1
({\cal O})$, where ${\cal O}$ runs over neighborhoods of ${\cal K}$ in ${\cal
R}^n$, and a compactum is said to be a carrier of $f\in S^{\prime 1}_\alpha$ if
$f$ has a continuous extension to the space assigned to it.

{\large Theorem 1}. {\it For any pair of compact sets ${\cal K}_1, {\cal K}_2
\subset {\cal R}^n$, the sequence

$$
0 \rightarrow S^{\prime 1}_{\alpha}({\cal K}_1\cap {\cal K}_2) \rightarrow
 S^{\prime 1}_{\alpha}({\cal K}_1) \oplus S^{\prime
1}_{\alpha}({\cal K}_2) \rightarrow S^{\prime 1}_{\alpha}({\cal
K}_1\cup {\cal K}_2) \rightarrow 0
         \eqno{(2)} $$

\noindent
is exact.}

For a proof, see \cite{14}. Its essential points are the same as those in the
paper of Kawai who treated the case of Fourier hyperfunctions $\alpha=1$ and
used, however, much more involved cohomology techniques. The mappings in
(2) are naturally defined via restrictions and the next to last arrow maps
a pair of hyperfunctions into the difference of their restrictions. This
theorem
expresses two simple but fundamental facts:

\begin{description}
\item[(i)] Every hyperfunction carried by ${\cal K}_1\cup {\cal K}_2$ can be
decomposed into a sum of two hyperfunctions with carriers ${\cal K}_1$ and
${\cal K}_2$.
\item[(ii)] If both ${\cal K}_1$ and ${\cal K}_2$ are carriers of a
hyperfunction, then so is ${\cal K}_1\cap {\cal K}_2$.
\end{description}
It is worthwhile to recall that in the case of tempered distributions and
ultradistributions the property (i) holds only for closed sets subject to some
regularity conditions. Furthermore from the statement (ii), by standard
compactness arguments, it follows that every hyperfunction has in ${\cal R}^n$
a
unique minimal carrier, the support.

In [13--14], we have argued that there is a reminiscence of these
properties in the nonlocalizable case. Namely, an analogue of Theorem 1 is
valid
for the closed cones with vertex at the origin if their associated spaces are
defined similarly, through the use of a filter of neighborhoods in  ${\cal
R}^n$. We refer to \cite{14,17} for terminology and notation concerning
cones, though these are quite customary.

{\large Definition 2}. {\it Let $\beta<1$ and let $K$ be a closed cone in ${\bf
R}^n$.  The space \sabk is defined to be the inductive limit of the spaces
$S_\alpha^{\beta}(U_\bullet)$, where $U_\bullet=U\cup B$, $U$ runs over open
cones
such that $K\subset \subset U$, and $B$ is the unit ball centered about the
origin.}

Strictly speaking, the ball must shrink to the origin, however this does not
change the space as may easily be verified by using the Taylor formula and as
follows, in particular, from Theorem 3 below. The only role of $B$ is to
provide
connectedness.

{\large Definition 3.} {\it A closed cone $K\subset {\bf R}^n$ is said to be a
carrier of  $f\in S^{\prime \beta}_\alpha$ if $f$ is continuous under the
topology induced on \sab by that of \sabk or, equivalently, if $f$  has a
continuous extension to \sabk.}

{\large Theorem 2}.  {\it For any $\beta <1$ and for any pair of closed cones
$K_1, K_2 \subset {\bf R}^n$, the sequence

$$
0 \rightarrow S^{\prime \beta}_{\alpha}( K_1\cap K_2) \rightarrow
 S^{\prime \beta}_{\alpha}(K_1) \oplus S^{\prime
\beta}_{\alpha}(K_2) \rightarrow S^{\prime \beta}_{\alpha}(K_1\cup
K_2) \rightarrow 0
         \eqno{(3)}$$

\noindent
is exact. Moreover it is topologically exact since all the involved spaces are
FS}.

This theorem formalizes the property that we call {\it angular localizability}.
We notice that, for a given $f\in S^{\prime
\beta}_\alpha$, it ensures the existence of a smallest carrier among the closed
cones which can be called the support cone of $f$. For $0<\beta<1$, Theorem 1
can be derived in an elementary way described in Sec.4. A more general, but
more
complicated proof covering the borderline case $\beta=0$ is presented in Sec.5.
We shall also show that \sab is dense in every space \sabk and so the
linear extension referred to in Definition 3 is unique. Actually, we shall
prove
the exactness of the dual sequence

$$
0 \leftarrow S^{\beta}_{\alpha}(K_1\cap K_2) \leftarrow
 S^{\beta}_{\alpha}(K_1) \oplus S^{\beta}_{\alpha}(K_2) \leftarrow
S^{\beta}_{\alpha}( K_1\cup K_2) \leftarrow 0 .
         \eqno{(4)}
$$

\noindent
This is completely equivalent to the initial problem since all the spaces here
are reflexive and their strong dual spaces are Fr\'echet, so the mappings in
(4) are of closed range if and only if the dual mappings possess this
property (see \cite{18}, Sec.IV.7.7) and our claim is proved by using the
formulae

$$
\hbox{Ker}\>u^\prime=(\hbox{Im}\>u)^\perp, \qquad
\overline{\hbox{Im}\>u^\prime}=(\hbox{Ker}\>u)^\perp
$$

\noindent
which is valid for any continuous linear mapping $u$  with dual $u^\prime$, and
wherein the bar stands
for the closure under the weak topology coinciding with that under the strong
topology for the reflexive spaces. Certainly the exactness of (4) in the term
$S^{\beta}_{\alpha}(K_1\cap K_2)$ is the only point that need be argued,
everything else being evident. In oher words, we have to prove that each
element
of $S^{\beta}_{\alpha}(K_1\cap K_2)$ can be decomposed into a sum of two
functions belonging to $S^{\beta}_{\alpha}(K_1)$  and
$S^{\beta}_{\alpha}(K_2)$.
As a first step, we shall represent the test function spaces under
consideration
in another, more convenient form.

\section{Representations of the test function spaces in terms of complex
variables}

{\large Theorem 3.} {\it Let $U$ be a nonempty open cone in ${\bf R}^n$ and let
$U_\bullet$ be as in {\rm Definition 2}. Denote by $d(x,U)$ the distance of $x$
from
$U_\bullet$ .  If $\beta<1$, the space   $S_\alpha^{\beta}(U_\bullet)$ is
identical to the
inductive limit of Banach spaces $E^{\beta,b}_{\alpha,a}(U)$ consisting of
entire analytic functions on ${\bf C}^n$ with the finite norms

$$
\Vert \varphi \Vert_{U,a,b}^\prime \> =\> \sup_{z}\vert \varphi (x)\vert
\exp\{-\rho_{{}_{U,a,b}}(z)\}, \eqno{(5)}
$$

\noindent
where

$$
\rho_{{}_{U,a,b}}(z)\>=\> -\vert x/a\vert^{1/\alpha} + d(bx,U)^{1/(1-\beta)} +
\vert by\vert^{1/(1-\beta)} .\eqno{(6)}
$$

\noindent
The same is true for empty $U$ if we set $d(x,U)=|x|$ and drop the first term
in} (6).

We remark that $d(bx,U)=bd(x,U)$ and  that in view of the inductive limit
procedure the choice of the norm in ${\bf R}^n$ is unessential here since all
these norms are equivalent. It should also be noted that for the particular
case
$U={\bf R}^n$ this reformulation is essentially due to Gelfand and Shilov, see
\cite{6}, Sec. IV.7.5.

{\it Proof}. We shall proceed along the same lines as in \cite{13}, where an
analogous representation has been obtained for
$S^\beta_\infty (U_\bullet)$. Due to the
condition $\beta<1$ the Taylor series expansion of $\varphi \in
S^{\beta,b}_{\alpha,a}(U_\bullet)$ is convergent for all $z \in {\bf C}^n$ and
since
any point $\xi \in U_\bullet$ can be taken as its center, the analytic
continuation
$\hat {\varphi}(z)$ is bounded by

$$
\Vert \varphi \Vert_{U,a,b}  \inf_{\xi \in U_\bullet} \inf_{k} \frac
 {a^{|k|}k^{\alpha k}}{|\xi^k|}\sum_{q} \frac{b^{|q|}q^{\beta q}}
{q!}|(z-\xi)^q| .\eqno{(7)} $$

\noindent
The infimum over k and sum over q can be estimated in the same manner as in
\cite{6} and are dominated by

$$
C\exp\{-|\xi/a'|^{1/\alpha} + |b'(x-\xi)|^{1/(1-\beta)} +|b'y|^{1/(1-\beta)}\}
$$

\noindent
with new constants $a',b'$. We see that, for any nonempty $U$, the space
$S_\alpha^{\beta}(U_\bullet)$ is trivial if $\alpha<1-\beta$, since then the
infimum
over $\xi$ is zero. Recall incidentally that $S^\beta_{1-\beta}$ is nontrivial
\cite{6} and this fact will be exploited in the next section. Taking  $\xi \in
U$ to be a point with minimal distance to $x$ and using the inequality

$$
-|\xi|^{1/\alpha}\>\leq \>-|x/2|^{1/\alpha} + |x-\xi|^{1/\alpha},
\eqno{(8)}
$$

\noindent
we infer that  $S_{\alpha,a}^{\beta,b}(U_\bullet)$ is continuously embedded
into a space $E^{\beta,b''}_{\alpha,a''}(U)$. In the degenerate case
$U=\emptyset$, the subscript is certainly inessential and we obtain the same
result with the above stipulation concerning $\rho$.

To prove the converse embedding, let $\varphi \in E^{\beta,b}_{\alpha,a}(U)$
and
let $\xi \in U_\bullet$. We employ Cauchy's inequality

$$
|\partial^q\varphi (\xi)|\>\leq\>q!r^{-q}\sup_{z\in D}|\varphi(\xi-z)|,
$$

\noindent
where $D\>=\>\{z\in {\bf C}^n: |z_j|\leq r_j, \quad j=1,...,n\}$, and use (5)
to
estimate $\varphi(\xi-z)$. Next we apply (8) with the interchanged position
of $\xi$ and $x$ and notice that the term $|x|^{1/\alpha}$ can be replaced by
$|x|^{1/(1-\beta)}$ if $\alpha\geq 1-\beta$. Then, since
$d(\xi-x,U)\>\leq\>d(\xi,U) + |x|\>\leq \>1 + |x|$ and since

$$
|x|^{1/(1-\beta)}  + |y|^{1/(1-\beta)}\>\leq \>h\sum_j |z_j|^{1/(1-\beta)}
$$

\noindent
for some $h>0$, we obtain

$$
|\partial^q\varphi (\xi)|\exp \{|\xi/2a|^{1/\alpha}\}\> \leq \> C\Vert \varphi
\Vert_{U,a,b}^\prime q!r^{-q}\exp \{\sum_j (b'r_j)^{1/(1-\beta)}\} .
$$

\noindent
The exponential on the left side can be replaced by $\sup_k
|\xi^k|/a^{\prime |k|}k^{\alpha k}$, and evaluation of the lower bound over $r$
yields the factor $b''^{|q|}q^{-(1-\beta)q}$. Thus we arrive at the inequality
$\Vert \varphi \Vert_{U,a',b''} \>\leq \>C\Vert \varphi \Vert_{U,a,b}^\prime$ .
The case of empty $U$ is treated with obvious simplifications and it remains to
say a little about the spaces $E^\alpha_\beta (U)=\hbox{inj\,lim}\,
E^{\beta,b}_{\alpha,a}(U)$ with $\alpha <1-\beta , U \neq \emptyset$. Let
$\varphi$ be an element of such a space. Evidently the second term in (6) is
negligible now and the function $\varphi(z) \varphi(iz)$ tends to zero as
$|z|\rightarrow \infty$. Hence by Liouville's theorem, it is identically zero
and the proof is finished.

{\large Theorem 4.} {\it For the entire analytic functions, the family of norms
{\rm (5)} is equivalent to that determined by the scalar products

$$
{\langle \varphi,\psi \rangle}_{U,a,b} = \int \bar{\varphi}(z)\psi
(z)\exp\{-2\rho_{{}_{U,a,b}}(z) \}{\rm d}v,
\eqno{(9)}
$$

\noindent
with ${\rm d}v$ denoting the Lebesgue measure on ${\bf C}^n$, and hence

$$
S_\alpha^{\beta}(U_\bullet) = {\rm inj\,lim}\,H^{\beta,b}_{\alpha,a}(U)
\quad  (a,b\rightarrow \infty),
\eqno{(10)}
$$

\noindent
where $H^{\beta,b}_{\alpha,a}(U)$ are the corresponding Hilbert spaces.

Proof}. Let $\Vert \cdot \Vert_{U,a,b}^{\prime  \prime}$ be the norm determined
by (9). Clearly $\Vert \varphi \Vert_{U,a,b}^{\prime  \prime}\leq C\Vert
\varphi \Vert_{U,a',b'}^\prime$ for $a>a', b>b'$. Conversely, by Cauchy's
integral formula,

$$
|\varphi (\zeta)|\> \leq \>C{\Vert \varphi \Vert}_{L^2 (D)}
\eqno{(11)}
$$

\noindent
for any polydisk $D$ centered about the point $\zeta$. One can multiply (11) by
$\exp\{-\rho_{{}_{U,a,b}}(\zeta) \}$, write $\zeta=z+(\zeta -z)$ on the right
side and then apply the triangle inequality to every term of the exponent
assuming that $|\zeta -z|\leq 1$. On doing so, we obtain

$$
|\varphi (\zeta)|^2\exp\{-2\rho_{{}_{U,a,b}}(\zeta) \}\> \leq \> C^\prime
\int\limits_{D}^{} |\varphi (z)|^2\exp\{-2\rho_{{}_{U,a',b'}}(z) \}{\rm d}v
$$

\noindent
with $a'<a,\, b'<b$ and hence ${\Vert \varphi \Vert}^\prime_{U,a,b}\leq
C^\prime
{\Vert \varphi \Vert}^{\prime \prime}_{U,a',b'}$, which proves the desired
identification (10).

\section{An elementary derivation of the density and decompositions theorems in
the case $\beta >0$.}

{\large Theorem 5.} {\it The space \sab, $0<\beta <1$, is dense in
$S_\alpha^{\beta}(U_\bullet)$ for any open cone $U\subset {\bf R}^n$, and all
the more
in each space \sabk, where $K$ is a closed cone. }

In \cite{14}, this result was obtained as a by-product of a theorem proven in
terms of real variables and expressing the angular-support property as a
fall-off property in the complementary directions. The representation \sabk =
inj lim$\, E^\beta_\alpha (U)$ enables one to present a simple direct proof.
Namely, let $\varphi \in  E^{\beta,b}_{\alpha,a} (U)$. We choose a function
$\chi_0 \in E^{\beta,b_0}_{1-\beta,a_0} $ so that $\int \chi_0 (\xi){\rm d}\xi
=
1$ and define an approximating sequence $\varphi_\nu$ by $\varphi_\nu (z)  =
\chi_\nu (z) \varphi (z)$, where $\chi_\nu$ is a sequence of Riemann sums for
the integral $\int \chi_0 (z-\xi){\rm d}\xi$ or, more explicitly,

$$
\chi_\nu (z) = \sum_{k\in {\bf Z}^n,|k|<\nu^2}\chi_0 (z-k/\nu )\nu^{-n}.
\eqno{(12)}
$$

\noindent
Clearly $\varphi_\nu \in S^\beta_{1-\beta} \subset S^\beta_\alpha $ if
$ a_0<1/b $. The series (12) converges uniformly on compact sets because it can
be dominated there by a convergent number series. The limit function is
analytic
and equals 1 at real points. Hence so does it on the whole of ${\bf C}^n$.
Furthermore the sequence $\varphi_\nu$ is bounded in the norm ${\Vert \cdot
\Vert}^\prime_{U,a,b'}$ with $b'=b+b_0$. By standard arguments, it follows that
${\Vert \varphi-\varphi_\nu \Vert}^\prime_{U,a',b''}\rightarrow 0$ for any $
a'>a, b''>b $ since

$$
\lim_{|z|\rightarrow \infty} \exp \{
\rho_{{}_{U,a,b'}}(z)-\rho_{{}_{U,a',b''}}(z)\} = 0 .
$$

\bigskip

{\large Theorem 6.} {\it Let $K_1$ and $K_2$ be closed cones in ${\bf R}^n$ and
let $K = K_1\cap K_2$. For every $\varphi \in S^\beta_\alpha (K) $, where
$0< \beta <1$, one can find a pair of functions $\varphi_j \in S^\beta_\alpha
(K_j) $ such that $ \varphi=\varphi_1+\varphi_2 $.

Proof}. We may assume that $\varphi \in E^{\beta,b}_{\alpha,a} (U) $, where $U$
is a cone-shaped neighborhood of $K$, and choose another open cone $W$ so that
$K\subset \subset W\subset \subset U$. Since the closed cones $K_j\backslash W$
have disjoint projections, there are open cones $W_j\supset \supset
K_j\backslash W_j $ with nonzero angular separation, that is,

$$
|x-\xi|\geq \theta |x|\quad \mbox{and} \quad |x-\xi|\geq \theta |\xi|\quad
\mbox{for all} \quad x \in W_1,\> \xi \in W_2,
\eqno{(13)}
$$

\noindent
where $\theta$ is a positive constant. Let us take $\chi_0 \in
E^{\beta,b_0}_{1-\beta,a_0} $  as before and set

$$
\chi(z) = \int \limits_{W_2}^{} \chi_0 (z-\xi){\rm d}\xi .
\eqno{(14)}
$$

\noindent
This function is entire analytic and we claim that one can define $\varphi_1$
to
be $\chi \varphi$. More specifically, if $a_0<\theta/b$, the product belongs to
the space $E^{\beta,b_1}_{\alpha,a} (U_1)$, where $K_1\subset \subset
U_1\subset
\subset W_1\cup U$ and $b_1$ is large enough. In fact, after taking $b_1\geq
b_0+b$, we need only to inspect the dependence on $x$. The inequalities (13)
show that, for $x\in W_1$, the bounded function $\chi (x)$ satisfies the
estimate

$$
|\chi(x)|\leq C_{a'}\exp\{-|\theta x/a'|^{1/(1-\beta)}\}
\eqno{(15)}
$$

\noindent
with any $a'>a$. Since $d(bx,U)\leq b|x|$, we see that $\chi \varphi$ decreases
inside $W_1\cup U$ no slower than $\varphi$ providing  $a_0<\theta/b$. Outside
this cone $d(bx,U_1)\geq \theta_1 |x|$ and therefore the additional
restriction $b_1>b/\theta_1$ ensures the bound

$$
|\varphi_1(z)|\leq C_1\exp \rho_{{}_{U_1,a,b_1}}(z)  ,
$$

\noindent
which proves the claim. On the other hand, $1-\chi$ satisfies an estimate
similar to (15) for $x$ in any $W_2^\prime \ss W_2$. Hence, by the same
arguments and under  condition $a_0<\theta^\prime/b$, the function
$(1-\chi)\varphi$ belongs to $E^\beta_\alpha (U_2)$, where $K_2\ss U_2\ss
W_2^\prime \cup U$. Thus

$$
\varphi = \chi \varphi + (1- \chi) \varphi
\eqno{(16)}
$$

\noindent
is the desired decomposition. This completes the proof of Theorem 5 and thereby
that of Theorem 2 for $\beta =0$.

\section{ General proof based on H\"{o}rmander's $L^2$--estimate}

Let us now turn to the case $\beta=0$. Because of triviality of $S_1^0$, the
above considerations fail to show the desired result. However a method
used for resolving the Cousin problem in the theory of analytic functions of
several complex variables points the way to get over this difficulty. Let
$\chi_0 (|x|)$ be any positive, smooth bump function with compact support in
the
unit ball.  Then functions constructed as above have the wanted behavior at
infinity but are nonanalytic. However one can put the decomposition (16) in
order by writing $\varphi_1 = \chi \varphi - \psi, \quad \varphi_2 = (1-\chi)
\varphi + \psi$, where $\psi$ must obey the nonhomogeneous Cauchy-Riemann
equations

$$
\partial \psi/\partial \bar{z}_j = \eta_j, \qquad j=1,\ldots, n
\eqno{(17)}
$$

\noindent
with $\eta_j = \varphi \partial \chi /\partial \bar{z}_j$. We notice that the
latter functions satisfy the compatibility conditions

$$
\partial \eta_j/\partial \bar{z}_k =\partial \eta_k/\partial \bar{z}_j,\quad
j,k=1,\ldots,n \eqno{(18)}
$$

\noindent
and have support in the 1-neighborhood of the boundary of $W_2$. By this
reason,
they vanish in the cone $W_1\cup W_2^\prime\cup U$ except for this
neighborhood,
where an decrease like $\exp\{-|x/a|^{1/\alpha}\}$ takes place. Therefore

$$
|\eta_j|\leq C_j \Vert \varphi \Vert_{U,a,b}^\prime  \exp
\rho_{{}_{U_1\cup U_2,a,b'}} ,
\eqno{(19)}$$

\noindent
where $U_1$ and $U_2$ are chosen as before  and $b^\prime$ is sufficiently
large.  In order to extend Theorem 6 to $\beta=0$, we only need to show that
the
system (17) has a solution with the same growth and fall-off properties as
those
of $\eta_j$'s. For this purpose, we shall take advantage of the representation
(10) and the following result of H\"{o}rmander \cite{19}.

Suppose that $\varrho \in \hbox{C}^2 ({\bf C}^n)$ is  a strictly \plu function,
i.e.,

$$
\kappa (z) = \inf_{\zeta} \sum_{j,k} \frac{\partial \varrho}{\partial z_j
\partial \bar{z}_k}\zeta_j \bar{\zeta_k}\left( \sum_j |\zeta_j|^2\right)^{-1} >
0.  $$

\noindent
Then for every collection of functions $\eta_j $ satisfying the
condition (18), one can find a solution $\psi $ of the system of equations (17)
such that $\psi \in L^2( {\bf C}^n, e^{-\varrho } {\rm d}v)$ and

$$
\int |\psi|^2 e^{-\varrho} {\rm d}v \leq \int |\eta|^2 e^{-\varrho} \kappa^{-1}
{\rm d}v . \eqno{(20)}
$$

\bigskip

{\large Lemma 1.} {\it There is a smooth \plu function $\rho^\ast_{{}_{U,a,b}}$
such that the replacement $\rho_{{}_{U,a,b}}\rightarrow
\rho^\ast_{{}_{U,a,b}}$ in the definition {\rm (5)} gives an equivalent family
of norms.

Proof}. Let us denote by $\rho_{\rm min}$ the greatest \plu minorant of
$\rho$. It exists for any continuous and even for upper semicontinuous function
\cite{17}. If $\varphi(z) $ is entire, then, modulus of an analytic function
being logarithmically \plu, the inequality $|\varphi|\leq C\exp \rho$ is
equivalent to $|\varphi|\leq C\exp \rho_{\rm min}$. The function
$\rho_{\rm min}$ is locally integrable but is not necessarily smooth. In
order to eliminate this trouble, one can exploit the regularization

$$
\rho^\ast (z) = \int \rho_{\rm min} (z-\zeta)\chi_0 (|\zeta|){\rm d}v
$$

\noindent
which is $\hbox{C}^\infty$ and \plu providing $\chi_0$ is nonnegative. As
before, by making use of the triangle inequality, we obtain

$$
\rho^\ast_{{}_{U,a,b}}(z) \leq \sup_{|\zeta|<1} \rho_{{}_{U,a,b}}(z-\zeta) \leq
\rho_{{}_{U,a',b'}}(z) + C  ,
$$

\noindent
where $a'>a,\, b'>b$ and can be taken arbitrarily close to $a,b$. It follows
that

$$
\rho^\ast_{{}_{U,a,b}}(z) \leq \rho^{\rm min}_{{}_{U,a',b'}}(z) + C .
$$

\noindent
Conversely, for any $\zeta$ in the unit ball, we have

$$
\rho^{\rm min}_{{}_{U,a,b}}(z)\leq \rho_{{}_{U,a',b'}}(z-\zeta) + C.
$$

\noindent
Since the left side of this inequality is \plu, one can replace the right
side by its greatest minorant and after that perform the multiplication by
$\chi_0 (\zeta)$ and the integration over $\zeta$, which gives

$$
\rho^{\rm min}_{{}_{U,a,b}}(z) \leq \rho^\ast_{{}_{U,a',b'}}(z) + C
$$

\noindent
and ends the proof of the lemma. Note that when $\beta=0$ and consequently
$\alpha>1$, the above inequalities are valid with $a'=a, b'=b$, that is, in
this case not only the whole family but also every norm is equivalent to that
obtained by the replacement $\rho\rightarrow \rho^\ast$.

Now we set

$$
\varrho(z) = 2\rho^\ast_{{}_{U_1\cup U_2,a,b}}(z) + 2\ln (1+|z|^2)
\eqno{(21)}$$

\noindent
with  $a$ and $b$ slightly greater than those in (19). Then $\kappa\geq
2(1+|z|^2)^{-2}$ and we are in a position to apply H\"{o}rmander's estimate,
which shows that the functions $\varphi_j$ corrected by adding $\psi$ have
finite ${\Vert \cdot\Vert}_{U,a_j,b_j}^{\prime \prime}$ -- norm, where $a_j$ is
arbitrarily close to the $a$ characterizing fall-off of the initial function
$\varphi$ and $b_j$ is large enough. On the other hand, it is well known that
each distribution and all the more any integrable function satisfying the
homogeneous Cauchy-Riemann  equations is actually a function analytic in the
usual sense.  Thus we have proved that Theorem 6 holds true for $\beta=0$.

In exactly the same manner one can extend to $\beta=0$ the density theorem.
Here
again we take $\chi_0$ with compact support and apply the former approximate
procedure (12). The sequence $\chi_\nu \varphi$ tends to $\varphi$ in the norm
$\Vert \cdot \Vert_{U,a,b}^{\prime \prime}$ with $a,b$ properly chosen but it
consists of nonanalytic functions. However $ \partial
\chi_\nu/\partial\bar{z}_j$ is uniformly convergent to zero and hence
${\Vert\varphi \partial \chi_\nu/\partial\bar{z}_j \Vert}_{U,a,b} \rightarrow
0$. This norm squared and doubled exceeds the integral on the right side of
(20)
if we set $\eta_j = \varphi \partial \chi_\nu/\partial\bar{z}_j $ and define
$\varrho (z)$ in the manner of (21).  Thus there exist functions $\psi_\nu$
such
that $\chi_\nu \varphi - \psi_\nu $ are analytic and compose a sequence
convergent to $\varphi$ in $S_\alpha^{\beta}(U_\bullet)$.

\section{Concluding remarks and outlook}

We consider the present work as a  step towards a distribution-theoretic
formalism for consistent treatment of gauge field models with arbitrarily
singular infrared behavior. We hope it may be of use in constructing nontrivial
field models on $S^0_\alpha({\bf R}^n)$ fulfilling all the Wightman axioms and
perhaps divergence-free at the cost of replacement of local commutativity by an
asymptotical commutativity condition. Besides this ultimate goal, we would like
to point out a few immediate applications of the obtained results. Firstly,
these lead to a natural generalization of the Paley-Wiener-Schwartz theorem to
the distributions defined on the space $S^0_\alpha$ and carried by a
closed cone. This in turn enables one to reformulate the generalized spectral
condition proposed by Moschella and Strocchi for infrared singular quantum
fields \cite{3} as a support  property of Wightman functions and then to
present
a more general formulation. Another possible application is the theory of
Lorentz invariant and Lorentz covariant distributions of arbitrarily high
singularity. In particular, this includes extension of the theorems
\cite{20,21}
concerning the structure of covariant tempered distributions, an invariant
splitting of distributions carried by the closed light cone, some theorems on
odd distributions, etc.

We conclude by noting that results similar to those listed above can be derived
for disributions on the space $S^0_\infty$ which is used in the nonlocal QFT
\cite{11,12}. In this case, a part of the derivations is even simpler since
$S^0_\infty$ is none other than the Fourier transform of Schwartz's space
${\cal D}$, i.e., such distributions are tempered in momentum space. However
the
topological structure of $S^0_\infty (K)$ is more complicated compared to
$S^0_\alpha (K)$, which gives rise to an additional trouble in proving an
analogue of the key theorem 6. One may overcome this difficulty in the same
manner as in showing the existence of support for $f\in S^{\prime 1}_\infty$ in
Refs. [14,15].

\bigskip

{\bf Acknowlegements}

\noindent
This work grew out of discussions with Professor V.~Ya.~Fainberg. I am grateful
for his continually helpful interest. I also thank the American Physical
Society for financial support.

\newpage

\xipt

\end{document}